\documentclass[conference]{IEEEtran}

\ifCLASSINFOpdf
\else
\fi
\usepackage{graphicx}
\usepackage{tabularx}
\usepackage{amsmath}
\usepackage[dvipsnames]{xcolor}
\usepackage{booktabs}
\usepackage{subcaption}
\usepackage{algorithm}
\usepackage{algpseudocode}
\begin{document}
%
\title{Video Popularity in Social Media: Impact of Emotions, Raw Features and Viewer Comments}



\author{\IEEEauthorblockN{Malika Ziyada ,
Pakizar Shamoi\IEEEauthorrefmark{1}}
\IEEEauthorblockA{School of Information Technology and Engineering \\
Kazakh-British Technical University\\
Almaty, Kazakhstan\\
Email: 
\IEEEauthorrefmark{1}p.shamoi@kbtu.kz
}
}


%


\maketitle



%
\IEEEpeerreviewmaketitle

\begin{abstract}


 The Internet has significantly affected the increase of social media users. Nowadays, informative content is presented along with entertainment on the web. Highlighting environmental issues on social networks is crucial, given their significance as major global problems. This study examines the popularity determinants for short environmental videos on social media, focusing on the comparative influence of raw video features and viewer engagement metrics. We collected a dataset of videos along with associated popularity metrics such as likes, views, shares, and comments per day. We also extracted video characteristics, including duration, text post length, emotional and sentiment analysis using the VADER and text2emotion models, and color palette brightness. Our analysis consisted of two main experiments: one evaluating the correlation between raw video features and popularity metrics and another assessing the impact of viewer comments and their sentiments and emotions on video popularity. We employed a ridge regression classifier with standard scaling to predict the popularity, categorizing videos as popular or not based on the median views and likes per day. The findings reveal that viewer comments and reactions (accuracy of 0.8) have a more substantial influence on video popularity compared to raw video features (accuracy of 0.67). Significant correlations include a positive relationship between the emotion of sadness in posts and the number of likes and negative correlations between sentiment scores, and both likes and shares. This research highlights the complex relationship between content features and public perception in shaping the popularity of environmental messages on social media.
 

\end{abstract}

\begin{IEEEkeywords}
environmental videos, social media, video popularity, video features, emotional analysis, sentiment analysis, popularity metrics, ridge regression, social media analytics
\end{IEEEkeywords}

\section{Introduction}
 
In the digital age, short social media videos can entertain and highlight significant environmental issues. The World Meteorological Organization’s 2023 report showed the highest global average surface temperature in a decade, reaching 1.45 $^{\circ}$Celsius \cite{wmoStateGlobal}. Thus, we should raise these issues in the Internet space. The reason is that the number of social network users is increasing yearly. For example, there are about 800 million users who have registered accounts on "Instagram", whereas "TikTok" attracted 500 million users during two years \cite{nazarov2021fuzzy}. 

As the number of online bloggers and organizations increases, so does the volume of online video content. For instance, videos could generate even one billion views in "TikTok" daily \cite{nazarov2021fuzzy}. However, the Internet offers both high-quality and low-quality videos to the audience. Bloggers being the voice of online community \cite{agarwal2014time} can even promote various social campaigns in social media \cite{khan2017modelling}. Thus, it's crucial to recognize the quality of trendy videos as they could heavily influence the audience and preferences. 

Furthermore, social media is one of the most effective tools for raising users' awareness about important problems \cite{peerj}, \cite{10453566}, \cite{nazzere}. Analyzing environmental content shows us how to make videos more engaging by improving sentiment and visuals. Promoting ecological content on social media attracts more attention and leads to effective solutions for these problems.

This study explores the correlation between raw video-specific features and their effect on media popularity in a specific environmental context. Moreover, we compare how user engagement in comments influences popularity. We achieve this by analyzing the gathered video dataset from the environmental Instagram account and defining popularity with the Ridge Regression Classifier by utilizing all necessary features.  We aim to answer the following questions:
\begin{itemize}
  \item What factors of video have the biggest impact on its popularity?
  \item How do the emotional features of video affect users?
    \item What affects video popularity more: the initial video parameters or public perception and audience engagement?
\end{itemize}

The main contributions of the paper are the following:
\begin{itemize}
 \item \textit{Dataset collection.} During the research, we gathered our own social media dataset with videos related to environmental issues. This is an important step because, unlike the public dataset, our dataset will provide insightful information. 
 We analyzed the audience activity on the Instagram social network. This enabled us to study the effectiveness of social media as a tool for solving significant problems.
\item \textit{Video features analysis. } We analyzed the impact of raw video features on popularity across various channels. For example, we studied how Instagram post descriptions, video duration, and brightness influence promotion.  
\item \textit{Classification task. } We used the ridge classifier to determine whether a video is popular.
\end{itemize}


 The paper has the following structure. Section I is an introduction. Section II presents earlier studies related to video popularity prediction. Section III includes an overview of methods, including data collection, video features, emotion analysis, and prediction model. Section IV provides the experimental results and main findings. Section V discusses limitations and ideas for future work. Finally, section VI provides concluding remarks.  

\section{Related Work}

This section provides an overview of related research in video popularity prediction on social media and video platforms. Researchers applied different methods to investigate the factors that impact the popularity of online content. Based on prior research, various methods, including regression, classification, neural networks, and ensemble techniques, can be used to identify the key factors of viral videos.

Numerous studies considered the application of classification algorithms as an effective method of prediction. For example, random forest, being the most accurate algorithm in \cite{ouyang2016peek}, considerably reduces the mean root square error errors of 32.25\% and 19.82\%  compared to the log-linear model and multi-linear model, respectively. Ouyang et al. \cite{ouyang2016peek} gathered data from the Chinese video platform "Youku" and analyzed various user and video features. These features included the video title, publish time, number of user followers, video duration, number of video tags, and others to predict future video view counts. Valet et al. also analyzed user engagement features such as likes, comments, and video views on the YouTube platform \cite{vallet2015characterizing}. As the previous researchers, they applied a classification approach, namely gradient boosted decision tree, which is considered as the common solution for the classification problems \cite{vallet2015characterizing}, \cite{friedman2001greedy}. With the help of this algorithm, they identified that YouTube features had more influence on the video popularity than the social media promotion posts related to the videos. 

Several studies described the regression algorithms used to predict how the video content became well-known. Authors in \cite{tan2018predicting}, \cite{lin2023tree}, \cite{chen2019social} also demonstrated that by identifying viral videos, it is possible to predict the top-N videos \cite{tan2018predicting} and even build the recommendation systems \cite{lin2023tree}. According to \cite{lin2023tree}, the tree-based progressive regression model, with a mean average error (MAE) of 4.741, demonstrates superior performance on the Kuaishou public dataset compared to weighted logistic regression, duration-deconfounded quantile, and ordinal regression, which result in MAE values of 6.047, 5.426 and 5.321 respectively. While \cite{lin2023tree} were more focused on the users’ videos’ watch time duration, other researchers in \cite{tan2018predicting}, \cite{chen2019social}, analyzed how the textual part impacts video promotion in social media. For example, authors applied multivariate linear regression to forecast the top-N videos \cite{tan2018predicting}. In contrast, the other study identified through XGBoost regression that integrating the textual and visual components positively impacts the popularity of the online content \cite{chen2019social}. 

Some works considered the neural networks in social media posts popularity prediction. While \cite{wu2017sequential}, \cite{ding2019social}, \cite{xu2020multimodal} analyzed either visual-language (V-L) or temporal (T) factors, authors in \cite{zhang2023improving} combined all of these intra-post (V-L) and inter-post (T) dependencies, including the category factor to build the Dependency-aware Sequence Network (DSN). By employing the newly proposed method, authors achieved the value of mean absolute error (MAE) of 1.192, outperforming the existing models such as Deep Context Neural Network (DTCN) \cite{wu2017sequential}, Multiple Layer Perceptron (MLP) \cite{ding2019social}, MLP with Attention Mechanism (Att-MLP) \cite{xu2020multimodal}, which produced values of 1.532, 1.483, 1.453, respectively. According to \cite{zhang2023improving} study, category factor plays a key role because it could predict whether the particular video category is viral. 

Next, the other recent study suggested using the ensemble learning algorithm to define the reputability of the videos on YouTube \cite{de2023classifying}. They classified videos according to their quality with the help of individual machine-learning algorithms and ensemble methods. The authors analyzed various attributes such as video views, likes, comments, duration, and other factors to determine whether the video would be considered reputable in their research. Consequently, the ensemble model resulted in a higher accuracy rate of 97.81\% for the training dataset, surpassing the best-performing individual algorithm, Random Forest, which achieved an accuracy of 97.01\% \cite{de2023classifying}. 

Research on video popularity in social media has identified several key factors. \cite{Jinna2017Multi} and \cite{Alexandros2021Video}  found that visual and social features, including emotional and color features, can predict image and video popularity in social media. \cite{wu2017sequential} and \cite{Raj2019Predicting} further emphasized the role of storytelling and emotional salience in increasing video popularity. Other studies highlighted the importance of early viewership and visual features in predicting video popularity \cite{F2014On}, \cite{Dalmoro2021Predicting}. A recent work \cite{Wang2022Two} demonstrated that combining emotional design features, such as color and anthropomorphism, can enhance learning outcomes in multimedia lessons.

Thus, video popularity prediction requires further research because the impact of each factor should be clearly defined. Moreover, as previous researchers mostly examined YouTube, Twitter, and public datasets, it is also important to consider another video platform—Instagram. As a result, by analyzing key factors influencing online videos, this study aims to identify how videos on Instagram could be promoted to raise important issues.

\section{Methods}

\begin{figure} [tb]
    \centering
    \graphicspath{{Images/}}
    \includegraphics [width = 0.4\textwidth]{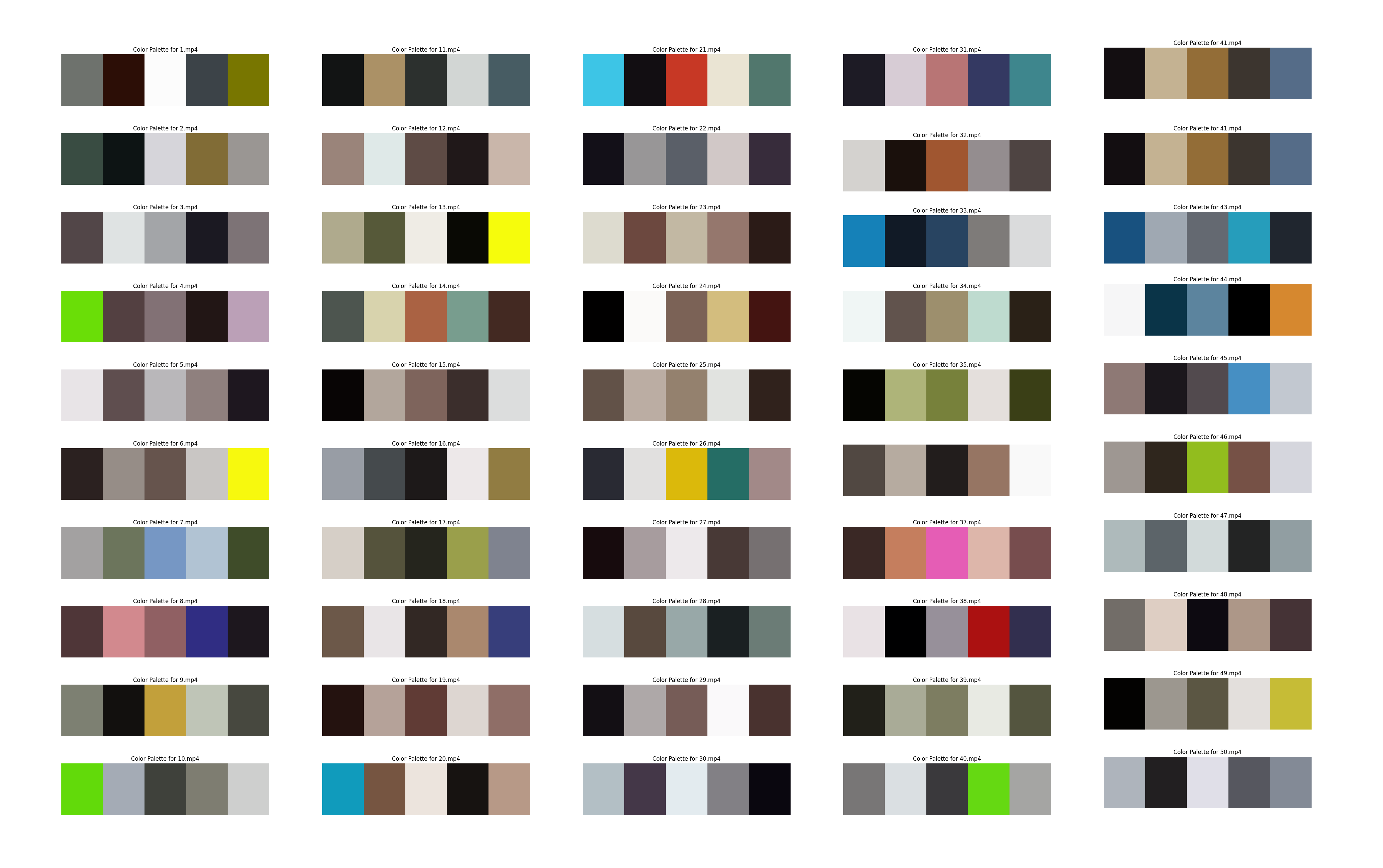}
    \caption{Color palettes of videos}
    \label{palettes}
\end{figure}

\begin{figure} [tb]
    \centering
    \graphicspath{{Images/}}
    \includegraphics [width = 0.4\textwidth]{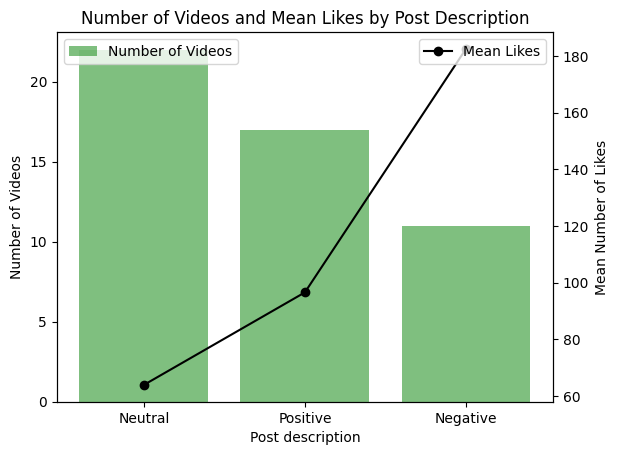}
    \caption{Post Description and Like Number}
    \label{descr_like}
\end{figure}

\begin{figure} [tb]
    \centering
    \graphicspath{{Images/}}
    \includegraphics [width = 0.4\textwidth]{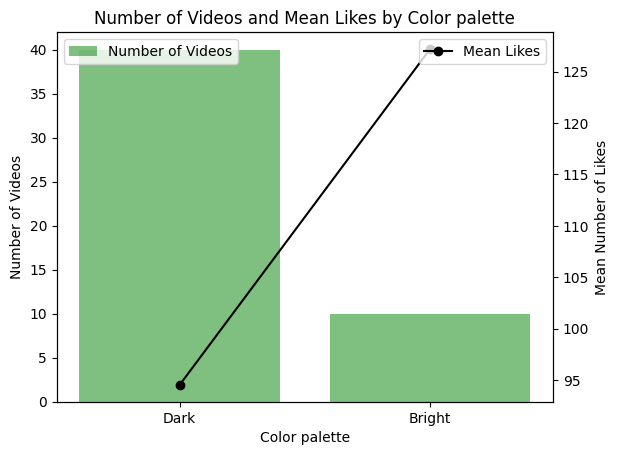}
     \caption{{Color Palette and Like Number}}
    \label{palet_like}
\end{figure}

\begin{figure*} [htb]
    \centering
    \graphicspath{{Images/}}
    \includegraphics [width =\textwidth]{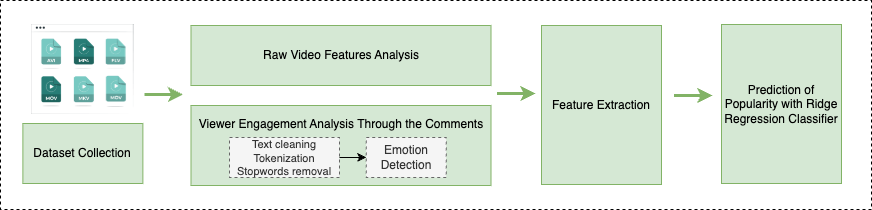}
    \caption{Methodology}
    \label{mainfig}
\end{figure*}

In this paper, we aim to apply the ridge classifier to identify the video popularity factors and their influence on users. The general scheme of the methodology can be seen in Fig. \ref{mainfig}. 

\subsection{Dataset Collection}

\subsubsection{Data Collection}
To begin with, we gathered a dataset of Instagram reel videos related to environmental issues. We collected 50 videos from an environmental account. To determine video popularity, we used input parameters such as "likes" and "views." We first calculated the threshold by computing the median values for these parameters. If a video's metrics exceed this threshold, it is considered popular; otherwise, it is non-popular.

\subsubsection{Dataset Description}
We retrieved the textual and visual information from the videos. Then, we included user engagement metrics such as likes, views, shares, and comments. We calculated each metric per day to avoid bias because of the upload date.

We analyzed the raw video features such as: 
\begin{itemize}
 \item \textit{Video Duration} 
 \item \textit{Length of the Post Text} 
\item \textit{Post Text Emotions}
\item \textit{Sentiment Score of the Post}
\item \textit{Brightness Level}
\end{itemize}

Moreover, we retrieved emotions from the users' comments to analyze the public perception. 

\begin{table*}[h]
\centering
\caption{Dataset of Raw Video Features }
\resizebox{\textwidth}{!}{%
\begin{tabular}{c|c|c|c|c|c|c|c|c|c|c|c|c|c|c}
\toprule 
\textbf{ID} & \textbf{Duration ( in sec.)}  & \textbf{Post Length} & \textbf{Post Emotion Happy} & \textbf{Post Emotion Angry} & \textbf{Post Emotion Surprise} & \textbf{Post Emotion Sad} & \textbf{Post Emotion Fear} & \textbf{Sentiment Score} &  \textbf{Average Intensity (Grayscale)} &\textbf{Likes} & \textbf{Views} & \textbf{Comments} & \textbf{Shared} & \textbf{Target}\\
\midrule 
1 & 7   & 65  & 0.0 & 1.0  & 0.0 & 0.0 & 0.0  & 0.0    & 94.79 & 83.30 & 2606.06 & 0.67 & 4.24 & 1 \\
2 & 85  & 288 & 0.0 & 0.0  & 0.0 & 0.0 & 1.0  & -0.318 & 94.17 & 70.05 & 1828.57 & 1.42 & 3.85 & 1 \\
3 & 180 & 376 & 0.0 & 0.0  & 0.11 & 0.22 & 0.67 & 0.8897 & 95.34 & 67.16 & 1621.62 & 1.41 & 2.81 & 0 \\
4 & 90  & 44  & 0.0 & 0.0  & 0.0 & 0.0 & 0.0  & 0.36 & 70.68 & 37.14 & 1619.61 & 0.30 & 0.30 & 0 \\
\bottomrule
\end{tabular}%
\label{table1}
}
\end{table*}

\begin{table*}[h]
\centering
\caption{Dataset of User's Emotions in Comments }
\resizebox{\textwidth}{!}{%
\begin{tabular}{c|c|c|c|c|c|c|c|c|c|c|c}
\toprule 
\textbf{ID} &
\textbf{Comment Emotion Happy} & \textbf{Comment Emotion Angry} & \textbf{Comment Emotion Surprise} & \textbf{Comment Emotion Sad} & \textbf{Comment Emotion Fear} & \textbf{Sentiment Score} & \textbf{Likes per day} & \textbf{Views per day} & \textbf{Comments per day} & \textbf{Shared per day} & \textbf{Target} \\
\midrule 
1 & 0.25 & 0.5 & 0.0 & 0.25 & 0.0 & -0.128 & 83.3030 & 2606.0606 & 0.6667 & 4.2424 & 1 \\
2 & 0.0 & 0.25 & 0.0 & 0.25 & 0.5 & -0.7556 & 70.0571 & 1828.5714 & 1.4286 & 3.8571 & 1 \\
3 & 0.56 & 0.0 & 0.44 & 0.0 & 0.0 & -0.7556 & 67.1622 & 1621.6216 & 1.4054 & 2.8108 & 0 \\
4 & 0.33 & 0.0 & 0.0 & 0.0 & 0.67 & -0.1546 & 37.1351 & 1621.6216 & 0.2973 & 0.2973 & 0 \\
\bottomrule
\end{tabular}%
\label{table2}
}
\end{table*}

\begin{figure} 
    \centering
    \graphicspath{{Images/}}
    \includegraphics [width = 0.5 \textwidth]{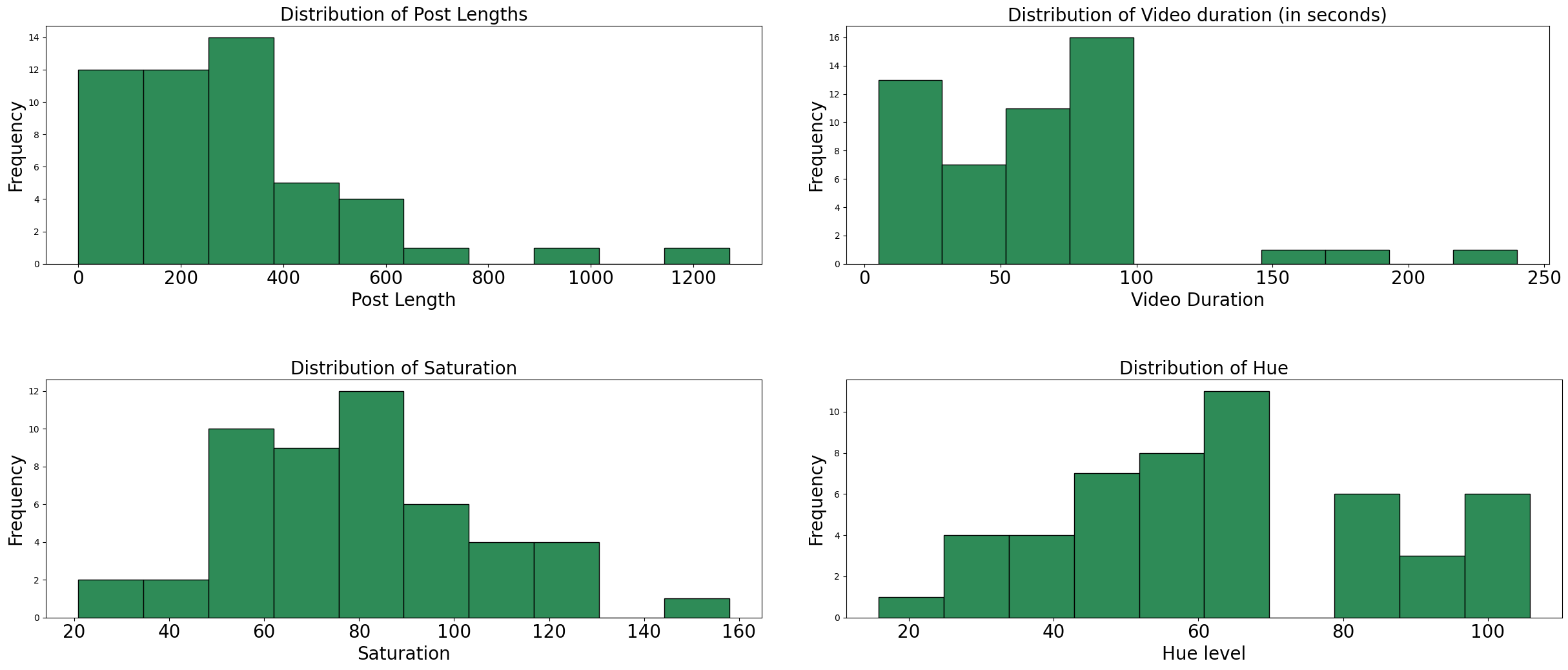}
    \caption{Video features distribution}
    \label{distr}
\end{figure}

\begin{figure} 
    \centering
    \graphicspath{{Images/}}
    \includegraphics [width = 0.5 \textwidth]{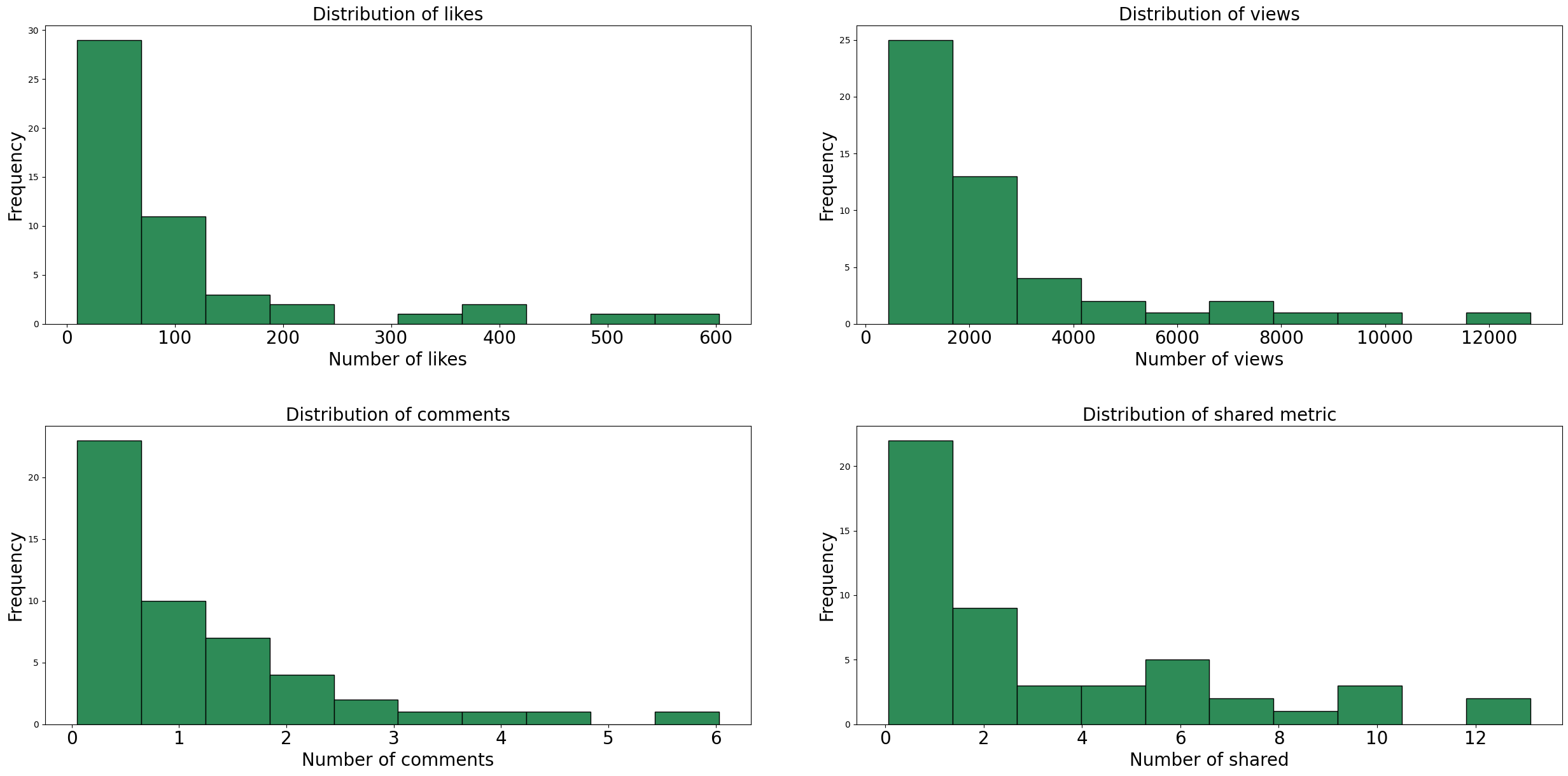}
    \caption{Popularity metrics distribution}
    \label{metrics}
\end{figure}

\subsection{Video Features Analysis}

To analyze video features, we focused on video duration and colors. We determined the dominant colors using the OpenCV-Python library {\cite{opencv} and K-means clustering. This process allowed us to extract each video's color palette and brightness intensity level. The extracted color palettes are in Fig. \ref{palettes}. Next, Fig. \ref{descr_like} and Fig. \ref{palet_like} demonstrate the relations between number of likes and post descriptions and color palettes, respectively.

 We presented the video features analysis results in Table \ref{table1} and Table \ref{table2}. Furthermore, we demonstrated video features and popularity metrics in Fig. \ref{distr} and Fig. \ref{metrics}, respectively. There is a concept of positivity bias stating that people prefer information that induces positive emotions \cite{michalos2014encyclopedia}. In contrast, our findings suggest that people like more negative posts.
 
\subsection{Sentiment and Emotion Analysis}
First, preparing the text data for further sentiment analysis was important.

\begin{itemize}
 \item \textit{Data Cleaning.} We removed special characters, links, and URLs from the post descriptions. 
 \item \textit{Tokenization.} It is better to break the sentences into smaller parts. Moreover, any irrelevant punctuation is also deleted from the general content. The split of the tokens leads to more efficient analysis. 
\item \textit{Stopwords.} To proceed with the further steps, we removed redundant words. For example, these words could include: "a", "the", "as", etc. Finally, we get the main content of the post.
\end{itemize}


Thus, we detected emotions through the post's content and its description. For this purpose, we used the text2emotion - Python package \cite{text2emotion}. By using this package, we got the five main emotions, such as fear, surprise, sadness, happiness, and anger, from the posts. Then, we complemented it with the sentiment score. We calculated the sentiment values with the SentimentIntensityAnalyzer supported by the Valence Aware Dictionary and Sentiment Reasoner (VADER). Figure \ref{metrics} shows the distribution of emotions across the text, where fear was common. 

Next, Fig. \ref{freq_word} represents the most frequent words from posts related to environmental issues. Moreover, according to the word cloud in Fig. \ref{fig:wordcloud_analysis}, the more dominant words are "people," "us," "climate," and "plastic." As a result, people should focus more on climate change and controlling plastic pollution.

\begin{figure} 
    \centering
    \graphicspath{{Images/}}
    \includegraphics [width = 0.4\textwidth]{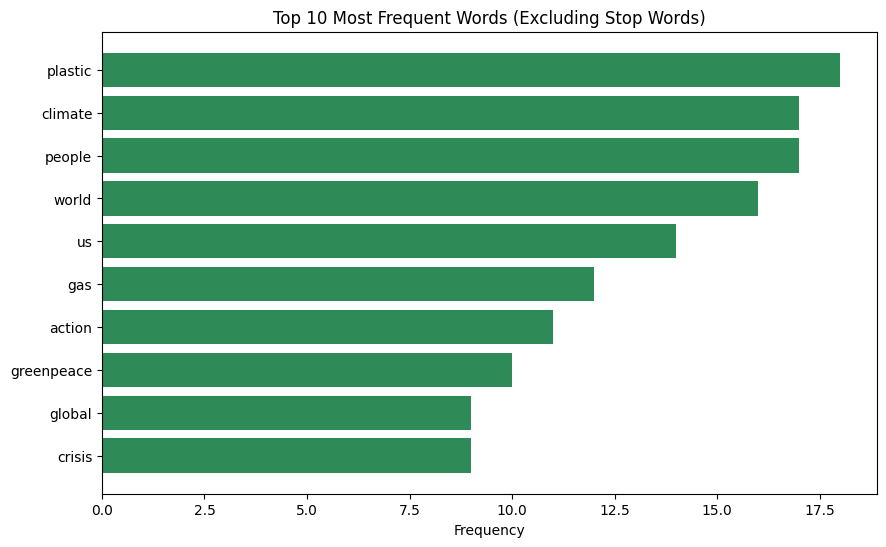}
    \caption{Frequent words}
    \label{freq_word}
\end{figure}

\begin{figure}[h!]
    \centering
    \graphicspath{{Images/}}
    
    \begin{subfigure}[b]{0.5\textwidth}
        \centering
        \includegraphics[width=\textwidth]{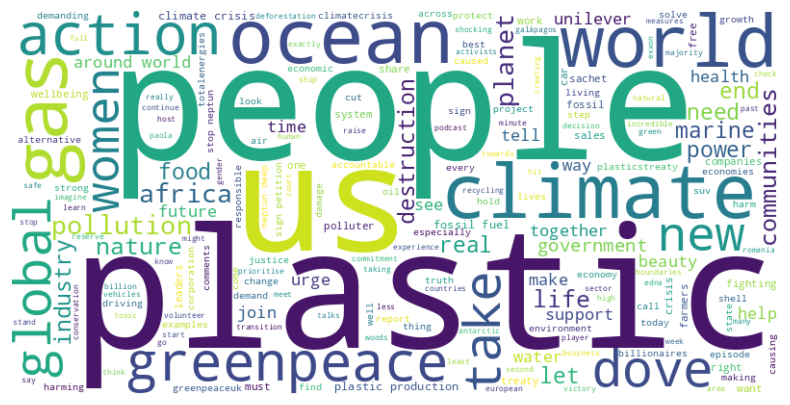}
        \caption{Word cloud for post text description}
        \label{fig:wordcloud_post_description}
    \end{subfigure}
    \hfill
    \begin{subfigure}[b]{0.5\textwidth}
        \centering
        \includegraphics[width=\textwidth]{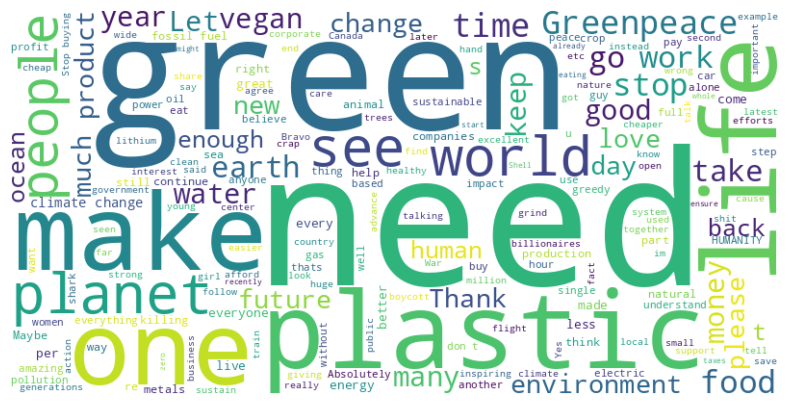}
        \caption{Word cloud for post comments}
        \label{fig:wordcloud_post_comments}
    \end{subfigure}
    
    \caption{Word cloud analysis for posts and comments}
    \label{fig:wordcloud_analysis}
\end{figure}

Fig. \ref{fig:emotions_analysis} presents the distribution of emotions in post descriptions and post comments. As we can see, the prevailing emotion is fear in both description and comments. This finding confirms the results of an earlier study \cite{10453566}.

\begin{figure}[h!]
    \centering
    \graphicspath{{Images/}}
    
    \begin{subfigure}[b]{0.5\textwidth}
        \centering
        \includegraphics[width=0.75\textwidth]{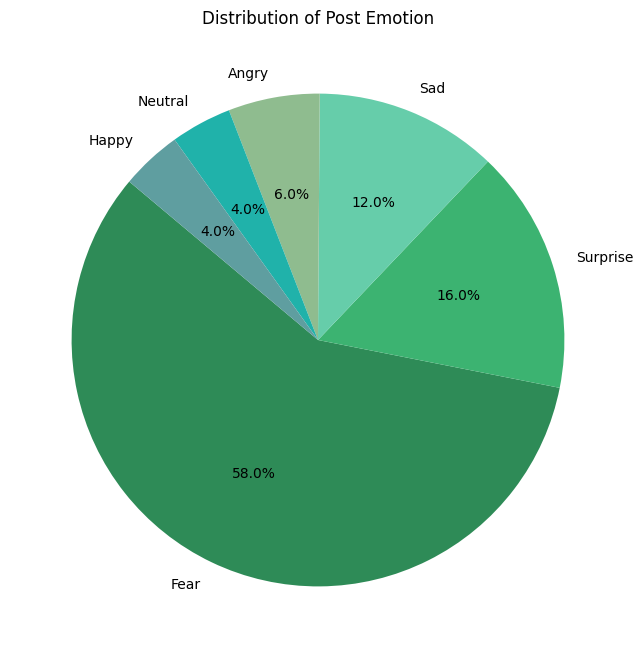}
        \caption{Post description emotions}
        \label{fig:post_description_emotions}
    \end{subfigure}
    \hfill
    \begin{subfigure}[b]{0.5\textwidth}
        \centering
        \includegraphics[width=0.75\textwidth]{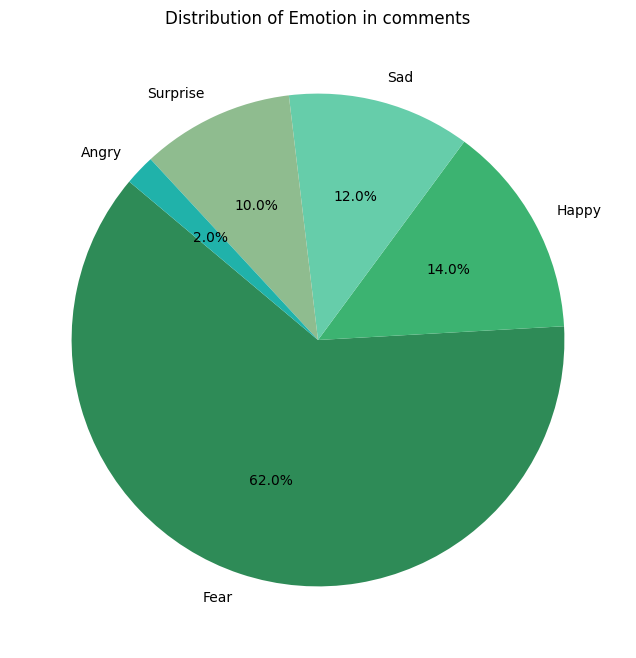}
        \caption{Post comments emotions distribution}
        \label{fig:post_comments_emotions}
    \end{subfigure}
    
    \caption{Emotions analysis in posts and comments}
    \label{fig:emotions_analysis}
\end{figure}

\subsection{Ridge Regression Classifier}
We utilized a Ridge Regression Classifier on the dataset. First, we split the data into training and testing sets, with a 0.7 and 0.3 ratio. Then, we evaluated the model's performance on the two datasets. Finally, we compared the effects of raw video features and user comments on popularity. Algorithm 1 presents the steps of the ridge regression classifier.

\begin{algorithm}
\caption{Ridge Classifier with Standard Scaler}
\begin{algorithmic}[1] 
\State \textbf{Input:} Training data $\{X_{\text{train}}, y_{\text{train}}\}$, Testing data $\{X_{\text{test}}, y_{\text{test}}\}$
\State \textbf{Parameters:} Regularization parameter $\alpha$

\State \textbf{Step 1: Standardize the Data}
\State \textbf{Fit Standard Scaler:}
\State \texttt{scaler $\gets$ StandardScaler()}
\State \texttt{scaler.fit}($X_{\text{train}}$)

\State \textbf{Transform Training Data:}
\State $X_{\text{train\_scaled}} \gets \texttt{scaler.transform}(X_{\text{train}})$

\State \textbf{Transform Testing Data:}
\State $X_{\text{test\_scaled}} \gets \texttt{scaler.transform}(X_{\text{test}})$

\State \textbf{Step 2: Initialize Ridge Classifier}
\State \texttt{ridge\_classifier $\gets$ RidgeClassifier(alpha=$\alpha$)}

\State \textbf{Step 3: Train Model}
\State \texttt{ridge\_classifier.fit}($X_{\text{train\_scaled}}$, $y_{\text{train}}$)

\State \textbf{Step 4: Predict on Test Data}
\State $y_{\text{pred}} \gets \texttt{ridge\_classifier.predict}(X_{\text{test\_scaled}})$

\State \textbf{Output:} Predictions $y_{\text{pred}}$
\end{algorithmic}
\end{algorithm}










\subsection{K-means Clustering}

We utilize k-means clustering \cite{kmeans1} \cite{kmeans2} across various color spaces to identify dominant colors from video frames. The procedure is outlined as follows \cite{aruzhan}:
\begin{itemize}
    \item Transform video frames from the standard RGB color space to a preferred model (e.g., HSV, CIELAB).
    \item Reformat the frame data into a 2D matrix where each row is a pixel and columns correspond to the color channels in the selected model.
    \item Implement k-means clustering on this matrix to segregate pixels into \( k \) groups, minimizing the within-cluster sum of squares:
    \[
    \min \sum_{i=1}^{k} \sum_{\mathbf{x} \in S_i} \|\mathbf{x} - \mathbf{m}_i\|^2
    \]
    where \( \mathbf{x} \) denotes a pixel, \( S_i \) is the cluster \( i \), and \( \mathbf{m}_i \) is the centroid of \( S_i \).
    \item The centroids \( \mathbf{m}_i \) effectively represent the dominant colors:
    \[
    \mathbf{m}_i = \frac{1}{|S_i|} \sum_{\mathbf{x} \in S_i} \mathbf{x}
    \]
    \item Convert these dominant colors back to RGB for visualization and analysis.
\end{itemize}
This method allows for the systematic extraction and comparison of dominant colors across different color models, facilitating a detailed analysis of color usage in environmental videos.

\section{Results}

\subsection{Correlation Heatmaps}

Fig. \ref{fig:rawfeatures} and Fig. \ref{fig:emotion_com} present the correlation heatmaps between popularity metrics and raw features, popularity metrics, and emotions in comments, respectively. As we can see, color palettes influenced popularity metrics. However, the emotional content of the posts and subsequent viewer reactions showed a stronger correlation with the popularity outcomes. There is a positive correlation between the emotion of sadness in posts and the number of likes, suggesting that videos evoking sadness tend to engage viewers more.
Negative correlations were observed between the sentiment scores of posts and both likes and shares, indicating that less positively perceived content might reduce viewer interaction. Next, surprise in posts was positively correlated with views, suggesting that unexpected content could increase viewer interest.

\begin{figure} 
    \centering
    \graphicspath{{Images/}}
    \includegraphics [width = 0.5\textwidth]{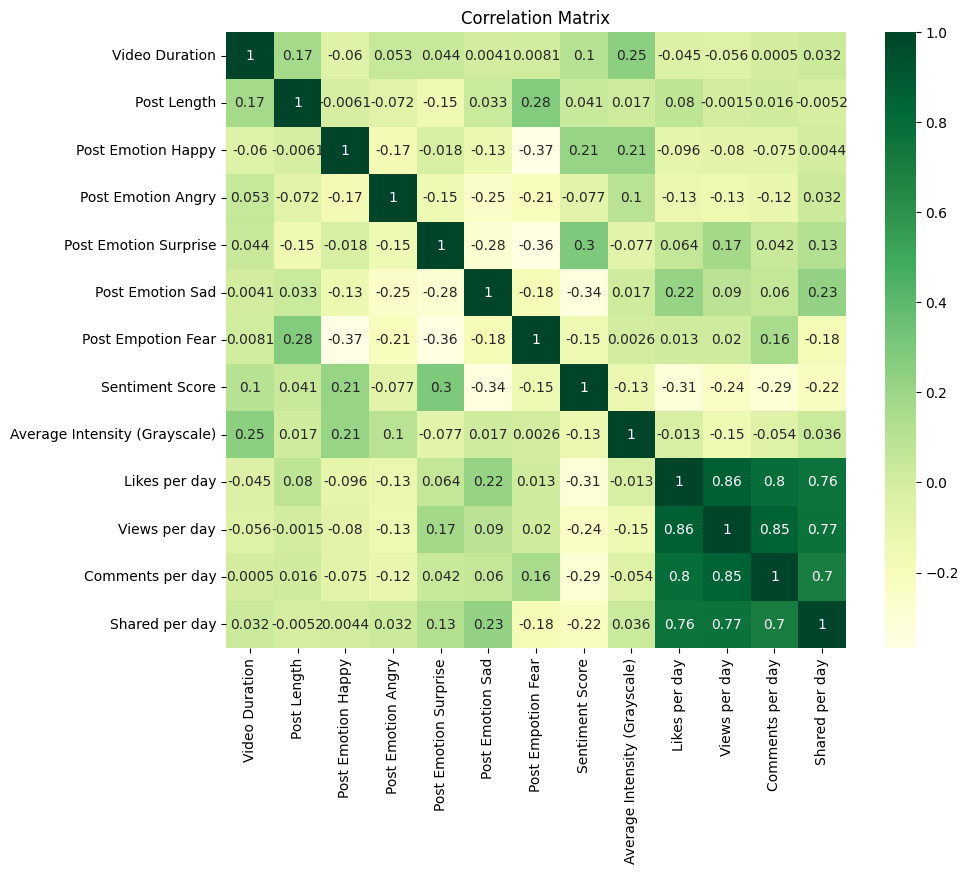}
    \caption{Raw Features vs. Popularity Metrics}
    \label{fig:rawfeatures}
\end{figure}

\begin{figure} 
    \centering
    \graphicspath{{Images/}}
    \includegraphics [width = 0.5\textwidth]{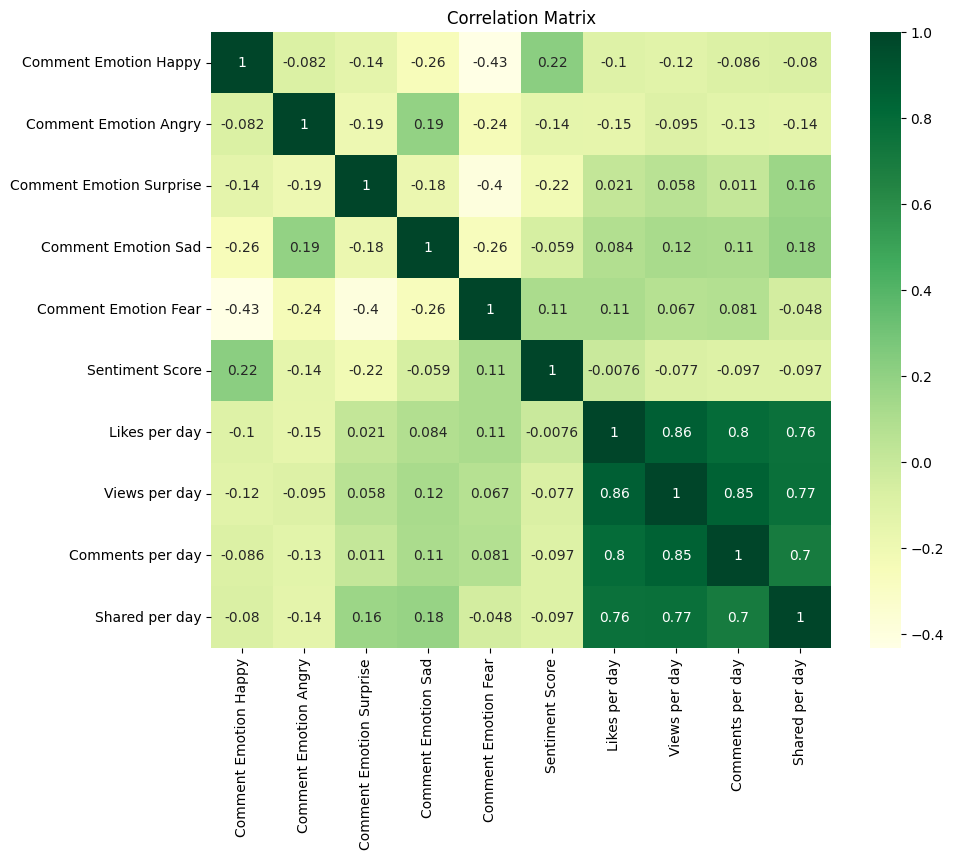}
    \caption{Emotion comments vs popularity}
    \label{fig:emotion_com}
\end{figure}

\subsection*{Evaluation Metrics}




Our research aims to classify videos as popular or not based on social media metrics. To assess the performance of our classification model, we use several standard statistical measures: Accuracy, Precision, Recall, and the \(F_1\) score.

\begin{itemize}
    \item \textbf{Accuracy} is the proportion of total correct predictions (both popular and not popular) made by the model relative to the total number of cases examined. It is given by the equation:
    \[
    \text{Accuracy} = \frac{\text{True Positives} + \text{True Negatives}}{\text{Total Number of Samples}}
    \]

    \item \textbf{Precision} reflects the accuracy of positive predictions. It measures the ratio of correctly predicted popular videos to all videos classified as popular by the model:
    \[
    \text{Precision} = \frac{\text{True Positives}}{\text{True Positives} + \text{False Positives}}
    \]

    \item \textbf{Recall}, also known as sensitivity, indicates the ability of the model to find all relevant instances of popular videos. It is calculated as:
    \[
    \text{Recall} = \frac{\text{True Positives}}{\text{True Positives} + \text{False Negatives}}
    \]

    \item The \(\textbf{F}_1 \textbf{ score}\) is the harmonic mean of Precision and Recall. This score is especially useful when the class distribution is imbalanced:
    \[
    F_1 \text{ score} = 2 \cdot \frac{\text{Precision} \times \text{Recall}}{\text{Precision} + \text{Recall}}
    \]
\end{itemize}


\subsection{Performance Evaluation}
 Table \ref{tab:results} present the specific classification metrics. Additionally, Table \ref{tab:results} demonstrates that popularity can be more accurately predicted based on audience activity and public opinion. However, the Comments Emotion Dataset outperforms the Raw Features Dataset regarding true negatives, false positives, accuracy, precision, and F1 Score. Specifically, it achieves higher accuracy (0.80 vs. 0.67) and precision (0.67 vs. 0.50), while recall remains consistent at 0.80 for both datasets.
This indicates that viewer engagement metrics after a posted video provide a more reliable basis for predicting video popularity than just using initial video attributes.



\begin{table}[tb]
\centering
\caption{Evaluation of Results}
\begin{tabular}{c|c|c}
\toprule
\textbf{Metric} & \textbf{Raw Features Data} & \textbf{Comments Emotion Data} \\
\midrule
Accuracy & 0.67 & 0.80 \\
Precision & 0.50 & 0.67 \\
Recall & 0.80 & 0.80 \\
F1 Score & 0.62 & 0.73 \\
\bottomrule
\end{tabular}
\label{tab:results}
\end{table}

\section{Limitations and Future Works}



This study has several limitations. First, using social media metrics as indicators of video popularity may not fully capture viewer sentiment due to potential manipulations such as paid promotions or bot-generated interactions, which may not reflect genuine user engagement. Additionally, the k-means algorithm, while effective for color extraction, may oversimplify the color dynamics within diverse video scenes. Future research could explore more advanced machine learning models to enhance the accuracy of popularity predictions. Also, our dataset is relatively small. Expanding the dataset to include a broader range of video content and incorporating longitudinal studies could provide deeper insights into trends.

For future research, employing more sophisticated machine learning techniques could improve the predictive accuracy of video popularity. Integrating dynamic modeling to track changes in viewer behavior and algorithm updates could also yield more robust insights. Additionally, examining a broader array of video types and incorporating qualitative content assessments could enhance our understanding of the factors influencing video popularity on social media. We also plan to gather more data and analyze more video features, such as subtitles, music, etc. We would also like to collect data from TikTok and YouTube. Next, we can determine which social media platform best highlights environmental issues.

\section{Conclusion}
This research investigated the impact of video and post characteristics on the popularity of environmental videos on social media. We discovered that viewer reactions post-publication significantly influence video popularity, demonstrating higher predictive accuracies (0.8) than initial video features alone (0.67). Our findings indicate that fear, frequently induced by video posts, is the predominant emotion, often leading to overall negative sentiments among viewers. These emotional responses, combined with factors like color and video duration, help explain variations in video popularity.
We found that most video posts and descriptions cause fear, which results in negative emotions among the audience. Moreover, after analyzing the text, we concluded that climate change and plastic pollution are significant problems today. 

Although the models used require further refinement to enhance their accuracy, the preliminary results are promising. The study findings can help ecology activists and governments effectively highlight environmental issues.






\section*{Acknowledgment}
This research has been funded by the Science Committee of the Ministry of Science and Higher Education of the Republic of Kazakhstan (Grant No. AP22786412)

\bibliography{mybib}
\end{document}